\documentclass[12pt,letterpaper,oneside]{article}
\usepackage{amssymb,tabularx,multirow,amsmath,natbib, epsfig, threeparttable, amstext, subfigure,xcolor}
\usepackage{pdflscape}
\usepackage{afterpage}
\usepackage{capt-of}
\usepackage{rotating}
\usepackage[makeroom]{cancel}

\newcommand{\blind}{0}

\newcommand{\rf}{\vskip .1in\par\sloppy\hangindent=1pc\hangafter=1
                 \noindent}

\def\spacingset#1{\renewcommand{\baselinestretch}%
{#1}\small\normalsize} \spacingset{1}

\usepackage[dvips,top=1.2in,bottom=0.65in,left=1.15in,right=1.15in,includefoot]{geometry}

\usepackage[margin=20pt,labelfont={bf,small},textfont={it,small},indention=0.5cm]{caption}

\begin{document}
\if0\blind
{
  \title{\bf Making Recursive Bayesian Inference Accessible}
  \author{Mevin B. Hooten\thanks{
    This research was funded by NSF EF 1241856 and NSF DMS 1614392.  Any use of trade, firm, or product names is for descriptive purposes only and does not imply endorsement by the U.S. Government.}\hspace{.2cm}\\
    U.S. Geological Survey \\ Colorado Cooperative Fish and Wildlife Research Unit \\ Department of Fish, Wildlife, and Conservation Biology \\ Department of Statistics \\ Colorado State University \\
    and \\
    Devin S. Johnson \\
    Marine Mammal Laboratory \\ Alaska Fisheries Science Center, NOAA Fisheries \\
    and \\
    Brian M. Brost \\
    Marine Mammal Laboratory \\ Alaska Fisheries Science Center, NOAA Fisheries} 
  \maketitle
} \fi

\if1\blind
{
  \bigskip
  \bigskip
  \bigskip
  \begin{center}
    {\LARGE\bf Making Recursive Bayesian Inference Accessible}
\end{center}
  \medskip
} \fi

\bigskip
\pagebreak
\begin{abstract}
Bayesian models provide recursive inference naturally because they can formally reconcile new data and existing scientific information.  However, popular use of Bayesian methods often avoids priors that are based on exact posterior distributions resulting from former studies.  Two existing Recursive Bayesian methods are: Prior- and Proposal-Recursive Bayes.  Prior-Recursive Bayes uses Bayesian updating, fitting models to partitions of data sequentially, and provides a way to accommodate new data as they become available using the posterior from the previous stage as the prior in the new stage based on the latest data.  Proposal-Recursive Bayes is intended for use with hierarchical Bayesian models and uses a set of transient priors in first stage independent analyses of the data partitions.  The second stage of Proposal-Recursive Bayes uses the posteriors from the first stage as proposals in an MCMC algorithm to fit the full model.  We combine Prior- and Proposal-Recursive concepts to fit any Bayesian model, and often with computational improvements.  We demonstrate our method with two case studies.  Our approach has implications for big data, streaming data, and optimal adaptive design situations.
\end{abstract}

\noindent%
\vfill

\newpage
\spacingset{1.45} 

\section{Introduction}
Bayesian methods have been incredibly useful for scientific inquiry because they empower the user to customize statistical analyses for their data and desired inference as well as formally incorporate existing scientific information (Gelman et al., 2012).  In particular, Bayesian hierarchical models (BHMs) also allow us to consider a complicated joint problem as a sequence of simpler conditional components.  In his seminal paper on BHMs, Berliner (1996) described the hierarchical model structure heuristically in terms of three quintessential components:  the data model, the process model, and the parameter model.  Each of these three components can be extended further, but the basic concept that statistical models should account for measurement error, process stochasticity, and parameter uncertainty, all simultaneously, is very powerful for making honest and reliable inference (Gelfand and Ghosh, 2015).  

Stochastic sampling approaches such as Markov chain Monte Carlo (MCMC; Gelfand and Smith, 1990) have facilitated the ability to fit a wide range of Bayesian models to data (Green et al., 2015).  However, as the size of data sets grow and the complexity of models increase, MCMC methods for fitting models have become limited in their applicability for big data settings (Brockwell, 2006).  Despite a proliferation of alternative sampling approaches (e.g., importance sampling, particle filtering, Hamiltonian Monte Carlo [HMC]; Doucet et al., 2001; Del Moral et al., 2006; Neal, 2011), MCMC is still popular, but also has fundamental weaknesses such as the inability to easily parallelize the computational procedure (beyond obtaining multiple chains; e.g., Glynn and Heidelberger, 1992; Bradford and Thomas, 1996; Rosenthal, 2000).                

Bayesian models also facilitate the formal use of preexisting information (resulting from former data analyses) in future data analyses.  However, despite widespread rhetoric claiming that previous Bayesian analyses can and should be incorporated into future data analyses as prior information, it is still rarely done in practice.  One potential hurdle to the formal incorporation of prior information is the inability to characterize the results of a previous data analysis as an analytically tractable prior with closed form (e.g., McCarthy and Masters, 2005; Garrard et al., 2012).  Thus, conventional practical guidance suggests approximating the joint posterior distribution resulting from a previous data analysis (using separate data) with an appropriate joint distribution (e.g., multivariate normal) and then use the approximate prior in the new data analysis.  This practice may yield a reasonable approximation in some cases, but is unsatisfying in the sense that recursive Bayesian analyses are not coupled exactly using well-accepted stochastic sampling methods such as sequential Monte Carlo (SMC), MCMC, and HMC to fit the models. 

In what follows, we discuss existing recursive Bayesian inference approaches and present a new method for performing recursive Bayesian inference using an advantageous combination of existing methods.  Our methods are helpful in a variety of situations, for both hierarchical and non-hierarchical Bayesian model fitting.  For ongoing data collection efforts, the procedure we describe allows us to represent previous data analyses as priors in new analyses.  We show that iterative inferential procedures can facilitate more rapid results using the methods we describe, especially when fitting the full model repetitively as new data arrives is infeasible.  Our approach can also be used to leverage parallel computing resources to accelerate the fitting of complicated Bayesian models such as those containing explicit dependence structure (e.g., Gaussian process models).  By partitioning data sets and applying the recursive Bayesian model fitting procedure, we show that our approach can lead to more efficient algorithms for fitting Bayesian models to big data sets.  Furthermore, our approach is accessible to practitioners and is compatible with SMC, MCMC, and HMC methods for fitting Bayesian models.  We demonstrate our methods with two case studies: a geostatistical model fit to environmental data and a hierarchical dynamic population model fit to ecological data.  

\section{Methods}
Also known as ``sequential'' inference or Bayesian filtering, recursive Bayes (RB) relies on fitting a statistical model to data in a series of steps (S\"arkk\"a, 2013).  Traditional RB inference has a natural appeal in studies where data are regularly collected over time and thus, it has been more commonly used in conjunction with state-space models (Chopin et al., 2013).  While the word ``sequential'' also appears in SMC, and SMC methods are relevant for RB (Chopin, 2002), they are not essential to the concept (as we describe in what follows).      
 
The general concept of performing an analysis in sequence is commonplace in many fields.  While many statistical methods are developed for analyzing a full data set in a single procedure, it may be advantageous to analyze data sets in groups.  For example, in addition to the situation where data arise sequentially, there may be computational advantages to analyze data in groups even when they are not indexed temporally.  In what follows, we review conventional RB based on a method we refer to as ``Prior-RB.''  We contrast Prior-RB with alternative recursive statistical procedures that rely on a sequence of stages meant to facilitate computation.  We refer to these approaches as ``Proposal-RB'' (for reasons that will become clear).  Finally, we combine these two recursive procedures to provide a framework for fitting Bayesian models more efficiently by leveraging parallel processing environments that are available in most modern computers.   

\subsection{Prior-Recursive Bayesian Inference}
Consider a generic data set $\mathbf{y}\equiv (y_1,\ldots,y_n)'$ and associated parametric statistical model $\mathbf{y} \sim [\mathbf{y}|\boldsymbol\theta]$, where $\boldsymbol\theta$ represents model parameters and we use the bracket notation `$[\cdot]$' to represent probability distributions (Gelfand and Smith, 1990).  For a specified prior $[\boldsymbol\theta]$, the posterior distribution is $[\boldsymbol\theta|\mathbf{y}]\propto [\mathbf{y}|\boldsymbol\theta][\boldsymbol\theta]$.  The main concept in Prior-RB is that, for a given partition of the data $\mathbf{y}\equiv (\mathbf{y}'_1,\mathbf{y}'_2)'$, we can find the posterior distribution associated with the first partition $[\boldsymbol\theta|\mathbf{y}_1]\propto [\mathbf{y}_1|\boldsymbol\theta][\boldsymbol\theta]$ and then use it as a prior in a secondary analysis of the second partition  

\begin{align}
  [\boldsymbol\theta|\mathbf{y}]&\propto [\mathbf{y}_2|\boldsymbol\theta,\mathbf{y}_1][\boldsymbol\theta|\mathbf{y}_1] \,, \\ 
  &\propto [\mathbf{y}_2|\boldsymbol\theta,\mathbf{y}_1][\mathbf{y}_1|\boldsymbol\theta][\boldsymbol\theta] \,.
\end{align}
The critical differences between the full model and the Prior-RB procedure are that 1) the second stage in the Prior-RB procedure requires knowledge of the conditional data model $[\mathbf{y}_2|\boldsymbol\theta,\mathbf{y}_1]$ and 2) the form of the posterior resulting from the first stage in Prior-RB $[\boldsymbol\theta|\mathbf{y}_1]$ must be known analytically.  However, if both distributions are known or at least well-approximated analytically, then we can make inference based on the full data set, but using only the second partition of data and the output from the first stage posterior.  This recursive concept is useful from a meta-analytic perspective because the same analyst does not have to compute the first stage posterior.  In fact, well-reported results of a previous data analysis based on a separate data set can serve as a sufficient statistic for reconciling inference based on both data sets.     

We can extend the basic concept of Prior-RB to accommodate multiple partitions of the data.  Suppose that we partition the data set into $J$ groups, $\mathbf{y}\equiv (\mathbf{y}'_1,\ldots,\mathbf{y}'_J)'$, then we can fit the first stage model as before to yield the posterior distribution $[\boldsymbol\theta | \mathbf{y}_1]$.  For the $j$th data partition, we obtain the posterior 

\begin{align}
  [\boldsymbol\theta|\mathbf{y}_{1:j}] &\propto [\mathbf{y}_j | \boldsymbol\theta,\mathbf{y}_{1:(j-1)}][\boldsymbol\theta|\mathbf{y}_{1:(j-1)}]  \,, \\  
  &\propto \left(\prod_{\nu=2}^j [\mathbf{y}_{\nu} | \boldsymbol\theta,\mathbf{y}_{1:(\nu-1)}] \right)[\mathbf{y}_1 | \boldsymbol\theta][\boldsymbol\theta] \,,  
\end{align}
\noindent where, $\mathbf{y}_{1:j}\equiv (\mathbf{y}'_1,\ldots,\mathbf{y}'_j)'$.  The $J$-partition Prior-RB procedure still requires analytical knowledge of each sequential posterior as well as the associated conditional data model $[\mathbf{y}_j | \boldsymbol\theta,\mathbf{y}_{1:(j-1)}]$ for $j=2,\ldots,J$.  

To illustrate the Prior-RB procedure, consider the binary data set $\mathbf{y}\equiv (0,1,1,1,0,0,0,1)'$, with data model $y_i \sim \text{Bern}(\theta)$ for $i=1,\ldots,n$ with $n=8$.  Based on a prior for $\theta$ specified as $\theta \sim \text{Beta}(1,1)$, the posterior is a classical result in Bayesian statistics:  $[\theta|\mathbf{y}]=\text{Beta}(\sum_{i=1}^n y_i + 1, \sum_{i=1}^n (1-y_i)+1)$, which is a beta distribution with both parameters equal to 5 in our example.  

To perform the Prior-RB method for this example with binary data, we split the data set into $J=4$ groups resulting in $\mathbf{y}_1\equiv (0,1)'$, $\mathbf{y}_2\equiv(1,1)'$, $\mathbf{y}_3\equiv (0,0)'$, and $\mathbf{y}_4\equiv (0,1)'$.  Then we analyze each data set recursively, using the appropriate conditional data model $[\mathbf{y}_j|\theta,\mathbf{y}_{1:(j-1)}]$ for each partition of data.  For this simple model, the conditional data model is $[\mathbf{y}_j|\theta]\equiv \text{Bern}(\theta)$ because the original data model assumed conditional independence of the data given $\theta$.   Thus, the Prior-RB method proceeds by finding each posterior recursively: $[\theta|\mathbf{y}_1]$, $[\theta|\mathbf{y}_{1:2}]$, $[\theta|\mathbf{y}_{1:3}]$, and $[\theta|\mathbf{y}_{1:4}]$.  It is easily shown that these are all beta distributions with parameter sets \{2,2\}, \{4,2\}, \{4,4\}, and \{5,5\}.  Thus, the Prior-RB method results in the same posterior distribution (i.e., $\text{Beta}(5,5)$) as fitting the model to all data simultaneously.  

The practical application of Prior-RB in settings involving more realistic statistical models and data sets involves two challenges: 1) The ability to find the required conditional data distributions and 2) the representation of the prior for the $j$th stage based on the $(j-1)$th stage posterior distribution.  These two challenges are exacerbated in the application of the Prior-RB method to situations where the data are not conditionally independent and/or more extensive hierarchical models are specified.  We revisit these issues in the sections that follow.  

\subsection{Proposal-Recursive Bayesian Inference}
When the data are not naturally ordered in time, it is not apparent how the Prior-RB concept may be helpful.  We address this idea in the following section, but first we set the stage for it by considering a slightly different form of recursive procedure to fit Bayesian models.  Suppose the model from the previous section is expanded to accommodate latent random effects $\boldsymbol\beta_j$ for $j=1,\ldots,J$ based on a natural partitioning of the data set $\mathbf{y}=(\mathbf{y}'_1,\ldots,\mathbf{y}'_J)'$ (not necessarily partitioned in time).  Then a generic hierarchical model structure for the data may be specified as

\begin{align}  
  \mathbf{y}_j &\sim [\mathbf{y}_j|\boldsymbol\beta_j] \,, \label{eq:hbm1} \\
  \boldsymbol\beta_j &\sim [\boldsymbol\beta_j|\boldsymbol\theta] \,, \label{eq:hbm2} \\
  \boldsymbol\theta &\sim [\boldsymbol\theta] \,, \label{eq:hbm3} 
\end{align}  
\noindent for $j=1,\ldots,J$ and where $\boldsymbol\beta_j$ are $p\times 1$ vectors and the data set partitions $\mathbf{y}_j$ are not necessarily equal-sized.  

For example, consider the situation where $J$ different data sets are collected by separate investigators and each set of coefficients $\boldsymbol\beta_j$ represent a subpopulation of interest.  Suppose that our main goal is to make population-level inference by characterizing the parameters $\boldsymbol\theta$.  These parameters ($\boldsymbol\theta$) give rise to the stochasticity associated with the subpopulation coefficients $\boldsymbol\beta_j$ and could represent, for example, an overall effect at the population level of a predictor variable on the response after accounting for subpopulation-level variation.  When the desired sample unit is the subpopulation, the hierarchical model in (\ref{eq:hbm1})--(\ref{eq:hbm3}) helps avoid pseudoreplication in the study (e.g., Hurlbert, 1984).   
 
The hierarchical model can also be thought of as a way to reconcile the results of separate data analyses in a meta-analysis framework.  Lunn et al.\ (2013) sought to use models with similar hierarchical structure as in (\ref{eq:hbm1})--(\ref{eq:hbm3}) to perform meta-analysis, synthesizing results across separate studies to obtain population-level inference for $\boldsymbol\theta$.  Assuming that each study used stochastic sampling methods (e.g., MCMC) to fit a Bayesian model to obtain a sample from the posterior distribution $[\boldsymbol\beta_j | \mathbf{y}_j]\propto [\mathbf{y}_j|\boldsymbol\beta_j][\boldsymbol\beta_j]$ based on the prior $[\boldsymbol\beta_j]$, Lunn et al.\ (2013) proposed a way to recursively use the results of these first stage analyses in a second stage to obtain population-level inference based on the full data set.  We refer to this approach as Proposal-RB because Lunn et al. (2013) suggested using the posterior samples from the subpopulation-level analyses as Metropolis-Hastings (M-H) proposals for $\boldsymbol\beta_j$ when fitting the full hierarchical model in (\ref{eq:hbm1})--(\ref{eq:hbm3}) using MCMC.       

The Proposal-RB approach is comprised of the following stages:  1) Specify subpopulation-level priors $[\boldsymbol\beta_j]$ and obtain a sample from the posterior distributions $[\boldsymbol\beta_j | \mathbf{y}_j]$ for $j=1,\ldots,J$ independently, then 2) fit the full model in (\ref{eq:hbm1})--(\ref{eq:hbm3}) using MCMC with M-H updates for $\boldsymbol\beta_j$ based on the previous stage posterior as a proposal ($\boldsymbol\beta_j^{(*)}$).  

The M-H acceptance probability for each $\boldsymbol\beta_j^{(*)}$ is $\text{min}(r_j,1)$ where 

\begin{align}
  r_j&=\frac{[\mathbf{y}_j | \boldsymbol\beta_j^{(*)}][\boldsymbol\beta_j^{(*)}|\boldsymbol\theta^{(k-1)}][\boldsymbol\beta_j^{(k-1)}|\mathbf{y}_j]}{[\mathbf{y}_j | \boldsymbol\beta_j^{(k-1)}][\boldsymbol\beta_j^{(k-1)}|\boldsymbol\theta^{(k-1)}][\boldsymbol\beta_j^{(*)}|\mathbf{y}_j]} \,, \\
   &=\frac{[\mathbf{y}_j | \boldsymbol\beta_j^{(*)}][\boldsymbol\beta_j^{(*)}|\boldsymbol\theta^{(k-1)}][\mathbf{y}_j | \boldsymbol\beta_j^{(k-1)}][\boldsymbol\beta_j^{(k-1)}]}{[\mathbf{y}_j | \boldsymbol\beta_j^{(k-1)}][\boldsymbol\beta_j^{(k-1)}|\boldsymbol\theta^{(k-1)}][\mathbf{y}_j | \boldsymbol\beta_j^{(*)}][\boldsymbol\beta_j^{(*)}]} \,, \\
   &=\frac{[\boldsymbol\beta_j^{(*)}|\boldsymbol\theta^{(k-1)}][\boldsymbol\beta_j^{(k-1)}]}{[\boldsymbol\beta_j^{(k-1)}|\boldsymbol\theta^{(k-1)}][\boldsymbol\beta_j^{(*)}]} \,, \label{eq:LunnMH} 
\end{align}
\noindent with $\boldsymbol\beta_j^{(*)}$ arising from the first stage posterior sample and MCMC iteration $k$ ($k=1,\ldots,K$).  The proposals $\boldsymbol\beta_j^{(*)}$ should be independent draws from the first stage posterior distribution for the cancellations to occur in the M-H ratio (\ref{eq:LunnMH}).  Thus, in practice, we sample $\boldsymbol\beta_j^{(*)}$ randomly (with replacement) from the first stage Markov chains to reduce autocorrelation (Lunn et al., 2013; Appendix A).  We then use a Gibbs, M-H, importance, or Hamiltonian update for the remaining model parameters $\boldsymbol\theta$ based on their full-conditional distribution $[\boldsymbol\theta|\cdot]\propto \left(\prod_{j=1}^J [\boldsymbol\beta_j|\boldsymbol\theta]\right)[\boldsymbol\theta]$ as usual (note that this full-conditional distribution does not involve $\mathbf{y}$).     

Benefits of the Proposal-RB suggested by Lunn et al.\ (2013) are that: 1) It provides a way to use output from a first stage analysis to fit a full hierarchical model where the first stage posterior distributions are well-represented, 2) it is not limited to meta-analysis, and 3) it can dramatically simplify the M-H ratio (\ref{eq:LunnMH}) because the data model cancels in the numerator and denominator.  Thus, using only output from $J$ independent model fits and knowledge of the first stage priors $[\boldsymbol\beta_j]$, we can fit the full model in (\ref{eq:hbm1})--(\ref{eq:hbm3}) to obtain inference.  

Aside from being a generally useful approach for fitting hierarchical models recursively, the Proposal-RB procedure is useful in data privacy situations where the original data cannot be released due to proprietary reasons, public safety, or legal restrictions (Altman, 2018) because the data do not appear in the second stage analysis.  Proposal-RB is also trivial to implement and is naturally adapted for parallel computing environments because we can sample from each of the transient posterior distributions $[\boldsymbol\beta_j|\mathbf{y}_j]$ in parallel at the first stage.      

To demonstrate the Proposal-RB approach, Hooten and Hefley (2019) fit a hierarchical Gaussian model to a set of eye region temperature data taken on a sample of 14 blue tits (\emph{Cyanistes caeruleus}).  These data arise from a study of individual-level versus population-level variation in wild birds by Jerem et al.\ (2018), who measured the eye region temperature (in Celsius) of a sample of blue tits using a non-invasive thermal imaging method.  For each individual, eye region temperature was recorded several times.  

Hooten and Hefley (2019) specified the full hierarchical Bayesian model for eye region temperature $y_{ij}$, for $j=1,\ldots,14$ and $i=1,\ldots,n_j$ repeated measurements per individual, as

\begin{align}
  y_{ij} \sim \text{N}(\mu_j,\sigma^2_j) \,, \\
  \mu_j \sim \text{N}(\mu,\sigma^2) \,, \\
  \sigma^2_j \sim \text{IG}(\alpha,\beta) \,, \\
  \mu \sim \text{N}(\mu_{0},\sigma^2_{0}) \,, \\
  \sigma^2 \sim \text{IG}(\alpha_{0},\beta_{0}) \,, 
\end{align}
\noindent where $\sigma^2_j$ are assumed to be individual-specific variance parameters in the model.  The random effects are the individual-level eye region temperature means $\mu_j$, and they are assumed to arise stochastically from a population-level distribution with mean $\mu$ and variance $\sigma^2$.  We used conventional MCMC to fit the full hierarchical model to the eye region temperature data (see Appendix B for implementation details and hyperparameters).    

We also fit the model in two stages using the Proposal-RB method.  In the first stage, we fit separate models at the individual-level resulting in MCMC samples from the posterior distributions $[\mu_j,\sigma^2_j|\mathbf{y}_j]\propto [\mathbf{y}_j|\mu_j,\sigma^2_j][\mu_j][\sigma^2_j]$ for $j=1,\ldots,J$.  For the transient priors, we specified $[\mu_j]\equiv \text{N}(\mu_\text{temp},\sigma^2_\text{temp})$ and $[\sigma^2_j]\equiv\text{IG}(\alpha, \beta)$, where $\mu_\text{temp}$, $\sigma^2_\text{temp}$ are treated as fixed hyperparameters in the first stage.  Using the MCMC output for $\mu_j$ and $\sigma^2_j$ from the first stage as proposals (Appendix A), we then fit the second stage model with M-H updates for $\mu_j$ and $\sigma^2_j$ jointly with an acceptance probability $\text{min}(r_j,1)$ for   

\begin{align}     
  r_j&=\frac{\prod_{i=1}^{n_j}[y_{ij}|\mu^{(*)}_j,\sigma^{2(*)}_j][\mu^{(*)}_j|\mu^{(k-1)},\sigma^{2(k)}][\sigma^{2(*)}_j]\prod_{i=1}^{n_j}[y_{ij}|\mu^{(k-1)}_j,\sigma^{2(k-1)}_j][\mu^{(k-1)}_j][\sigma^{2(k-1)}_j]}{\prod_{i=1}^{n_j}[y_{ij}|\mu^{(k-1)}_j,\sigma^{2(k-1)}_j][\mu^{(k-1)}_j|\mu^{(k-1)},\sigma^{2(k)}][\sigma^{2(k-1)}_j]\prod_{i=1}^{n_j}[y_{ij}|\mu^{(*)}_j,\sigma^{2(*)}_j][\mu^{(*)}_j][\sigma^{2(*)}_j]} \;, \\
  &=\frac{[\mu^{(*)}_j|\mu^{(k-1)},\sigma^{2(k)}][\mu^{(k-1)}_j]}{[\mu^{(k-1)}_j|\mu^{(k-1)},\sigma^{2(k)}][\mu^{(*)}_j]} \,. \label{eq:Lunn_MH_temp}
\end{align}  

The resulting M-H ratio in (\ref{eq:Lunn_MH_temp}) simplifies to a ratio of the conditional random effect distributions and transient priors from the first stage and does not include the priors for $\sigma^2_j$ because they cancel.  We used standard Gibbs updates for the remaining population-level parameters $\mu$ and $\sigma^2$ because their full-conditional distributions are both conjugate (Appendix B).  

\begin{figure}[htp]
  \centering
  \includegraphics[width=6in]{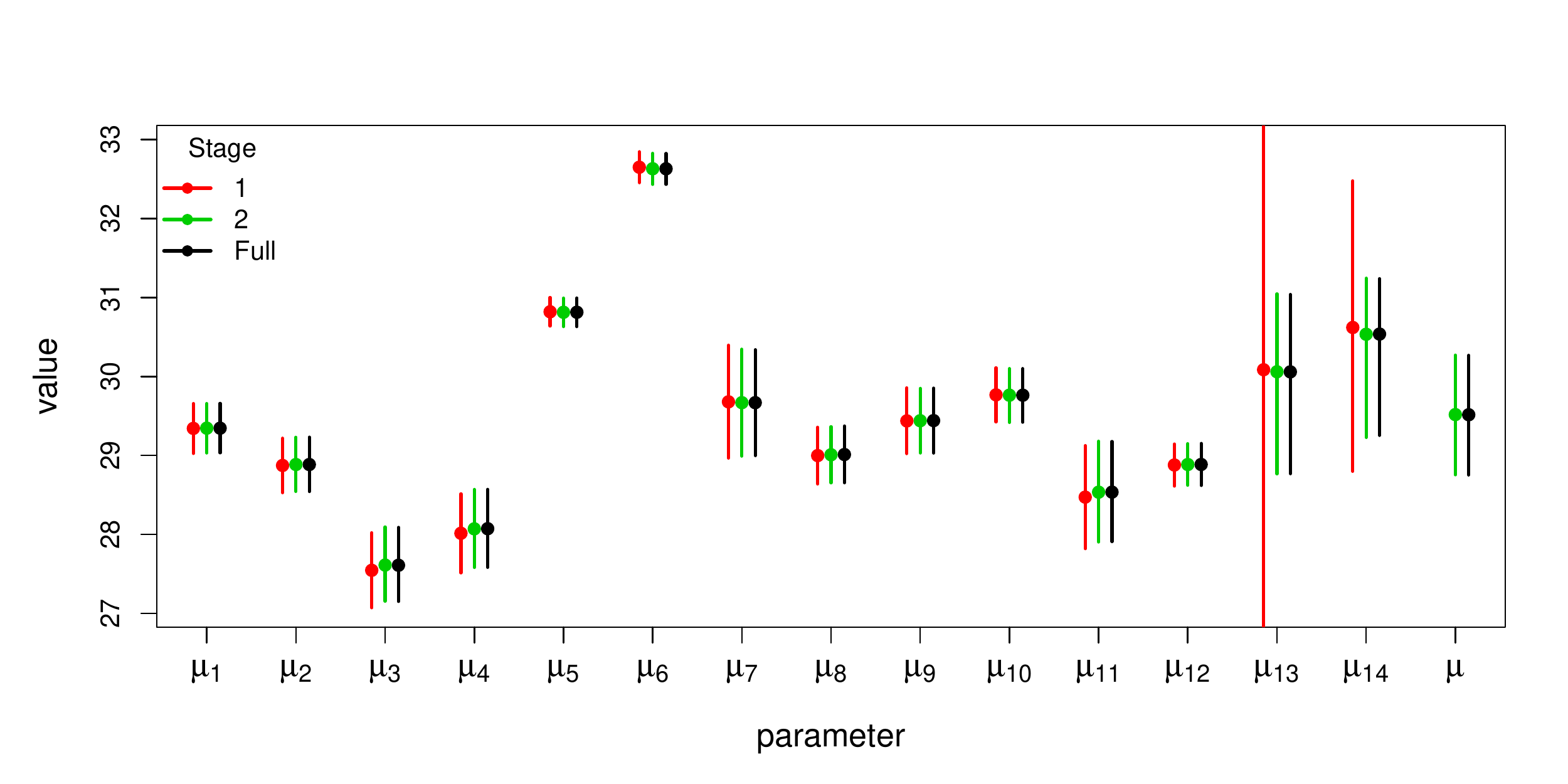}
  \caption{Marginal posterior means (points) and 95\% credible intervals (vertical lines) for the individual-level ($\mu_j$) and population-level ($\mu$) means.  The black intervals correspond to the results from the full model and the red and green intervals correspond to the results from the first- and second-stage analysis, respectively.}
  \label{fig:hier_L_post}
\end{figure}

We compared the marginal posterior means and 95\% credible intervals for the individual-level eye region temperature parameters $\mu_j$ and the population-level mean $\mu$ in Figure~\ref{fig:hier_L_post}.
The results of the first stage analysis, which did not account for the natural regularization induced by the hierarchical model structure, indicate larger posterior means and credible intervals (red bars in Figure~\ref{fig:hier_L_post}).  In contrast, the results of the second stage analysis (green bars in Figure~\ref{fig:hier_L_post}) match those from fitting the full model to the entire data set simultaneously (black bars in Figure~\ref{fig:hier_L_post}).  In this example, the individual-level inference is nearly identical for individuals 1--12, where sample sizes ranged from 10 to 60 individual-level measurements.  However, for individuals 13--14, there were only two measurements each, resulting in more uncertainty in first stage analysis for those effects.  Nonetheless, by shrinking the individual-level parameters toward the population-level mean in the second stage of the Proposal-RB procedure, we obtain the correct inference.   

There may not be a computational advantage when using Proposal-RB to fit the hierarchical model to the blue tit eye region temperature data because the model is fairly simple and the data set is small.  However, if the first stage models were fit by separate investigators, one would not need the original data to perform the population-level meta-analysis and make inference on $\mu$ and $\sigma^2$.  

When the models and/or data sets are more complex, the Proposal-RB method can lead to computational improvements.  For example, Hooten et al.\ (2016) and Gerber et al.\ (2018) applied the Proposal-RB method to make population-level inference about resource selection by animals.  In particular, Hooten et al.\ (2016) showed that the Proposal-RB approach suggested by Lunn et al.\ (2013) may be particularly useful for cases where the data model is numerically challenging to evaluate.  Specifically, Hooten et al. (2016) considered a hierarchical point process model for animal telemetry data where the data model was specified as a product of weighted distributions (Patil et al., 1977) that contained integrals of a function that included model parameters.  These integrals were a crux in implementing the spatial point process model because an optimization or stochastic sampling algorithm such as MCMC must numerically calculate the integral repeatedly when fitting the model to data.  The Proposal-RB approach used by Hooten et al.\ (2016) simplified the sampling procedure in the second stage analysis substantially because the integrals in the data model did not need to be calculated again after the first stage, resulting in a procedure that required less computational time than fitting the full model jointly.  

Overall, it is clear that the Proposal-RB method is useful for fitting certain classes of hierarchical models to data that are naturally partitioned.  However, Proposal-RB does not directly translate to non-hierarchical models and cases where the data are not conditionally independent.  When the data are not conditionally independent, we can still fit independent models for each partition in the first stage, but the data model will not cancel in the second stage M-H updates for $\boldsymbol\beta_j$.  If the data model is trivial to calculate, it is possible that the Proposal-RB approach may still be useful, but data models with dependence (e.g., Gaussian process models) can be numerically challenging to calculate repetitively.        

Similarly, for non-hierarchical models, natural partitions of the data may not exist and it becomes more difficult to envision useful partition-specific first stage models.  While it may be possible to contrive an auxiliary variable approach that augments a non-hierarchical model with an artificial latent process (e.g., Albert and Chib, 1993), we propose a simpler alternative in what follows.     

\subsection{Prior-Proposal-Recursive Bayesian Inference}
Proposal-RB is useful for meta-analysis and fitting hierarchical models, but the concepts in Proposal-RB do not automatically transfer to non-hierarchical models, or if so, may not be helpful computationally.  Therefore, we propose a combination of Prior- and Proposal-RB (hereafter, PP-RB) concepts that makes RB more accessible to practitioners and facilitates inference for model parameters for a wide class of Bayesian models.  

Our PP-RB approach assumes the data can be partitioned as described earlier such that $\mathbf{y}\equiv (\mathbf{y}'_1,\ldots,\mathbf{y}'_J)'$ and we can implement the Prior-RB procedure for recursively fitting the full model in stages.  To implement the PP-RB approach, we first obtain a sample from $[\boldsymbol\theta | \mathbf{y}_1]$ as before, then, for the next $J-1$ stages, we recursively obtain samples from            
\begin{equation}
  [\boldsymbol\theta|\mathbf{y}_{1:j}] \propto [\mathbf{y}_j | \boldsymbol\theta,\mathbf{y}_{1:(j-1)}][\boldsymbol\theta|\mathbf{y}_{1:(j-1)}] \,, \label{eq:PP-RBI_post}
\end{equation}
\noindent for $j=2,\ldots,J$.  Borrowing the technique from Proposal-RB where we use the transient posterior from the previous stage as the proposal (in addition to the prior, as in Prior-RB), our M-H acceptance probability for the $j$th stage and $k$th MCMC iteration can be written as $\text{min}(r_j,1)$ with 

\begin{align}
  r_j&=\frac{[\mathbf{y}_j | \boldsymbol\theta^{(*)},\mathbf{y}_{1:(j-1)}][\boldsymbol\theta^{(*)}|\mathbf{y}_{1:(j-1)}][\boldsymbol\theta^{(k-1)}|\mathbf{y}_{1:(j-1)}]}{[\mathbf{y}_j | \boldsymbol\theta^{(k-1)},\mathbf{y}_{1:(j-1)}][\boldsymbol\theta^{(k-1)}|\mathbf{y}_{1:(j-1)}][\boldsymbol\theta^{(*)}|\mathbf{y}_{1:(j-1)}]} \,, \\
     &=\frac{[\mathbf{y}_j | \boldsymbol\theta^{(*)},\mathbf{y}_{1:(j-1)}]}{[\mathbf{y}_j | \boldsymbol\theta^{(k-1)},\mathbf{y}_{1:(j-1)}]} \,, \label{eq:PP-RBI_MH} 
\end{align}
\noindent where $\boldsymbol\theta^{(*)}$ is the $k$th realization from the transient posterior sample from the previous stage.  Notice that the M-H ratio in (\ref{eq:PP-RBI_MH}) consists only of a ratio of the conditional data models.  Thus, the PP-RB approach still requires the knowledge and calculation of the conditional data model at each MCMC iteration and stage.  However, because the set of posterior realizations we use as proposals throughout the procedure are acquired as a result of the first stage analysis, we can pre-calculate the log density (or mass function) of the conditional data model for each proposal $\boldsymbol\theta^{(*)}$ in parallel between stages 1 and 2 in the procedure.  With values for the numerator in the M-H ratio resulting from our quasi-prefetching technique (i.e., the pre-calculation of log densities for all possible proposals of $\boldsymbol\theta$; Brockwell, 2006), performing the updates for $\boldsymbol\theta$ is less computationally intensive.  Furthermore, because we need only save the values for $\log[\mathbf{y}_j | \boldsymbol\theta^{(*)},\mathbf{y}_{1:(j-1)}]$ after the first stage, the PP-RB approach has low memory requirements between stages.     

\section{PP-RB Application to Geostatistics}
\begin{figure}[htp]
\begin{center}
\includegraphics[width=.8\textwidth]{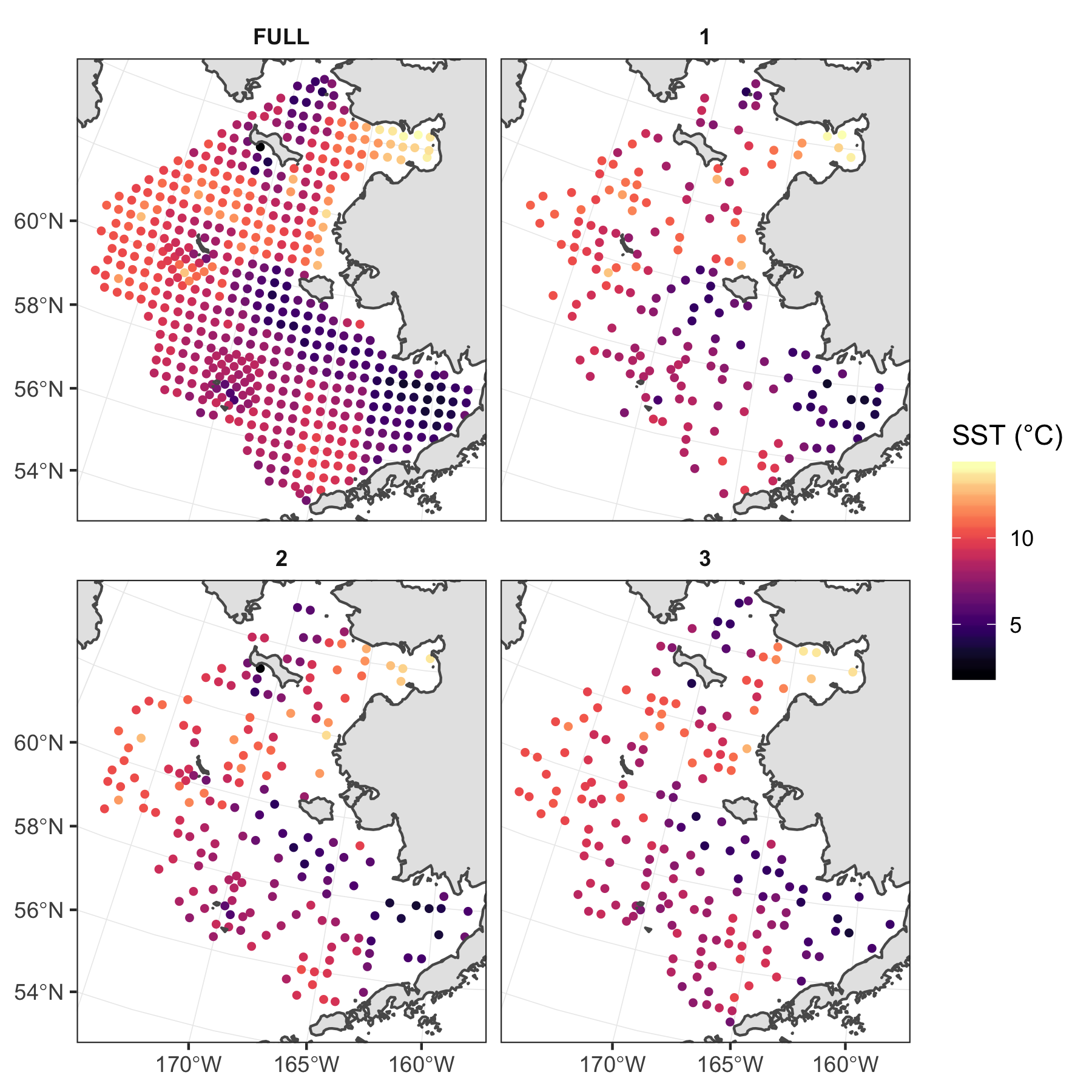}
\end{center}
\caption{\label{fig:geo_sst} Sea surface temperature measurements ($^o$C) from NOAA Fisheries 2017 bottom trawl survey on the eastern and northern Bering Sea shelf off the coast of Alaska. The `FULL' plot depicts all data together, while the remaining plots show the locations used in each partition of the data for the PP-RB approach.}
\end{figure}

Our PP-RB approach can be applied to fit a wide range of Bayesian models recursively.  As a first demonstration of the PP-RB method, we apply it to fit the standard geostatistical model (Cressie, 1990), which is very commonly used in environmental and ecological applications. The data used for this illustration are measurements of sea surface temperature (SST) on the eastern and northern Bering Sea shelf near the Pribilof Islands, Alaska. The data were obtained as part of the 2017 NOAA Fisheries bottom trawl surveys used to assess the condition of groundfish stocks in the Bering Sea. The SST measurements are  collected in the same locations as the fishing trawls (Figure \ref{fig:geo_sst}, `FULL'). There are $n=520$ observations in this data set spaced on a 20km grid with additional locations surveyed near the Pribilof Islands and St. Matthew Island.

Parametric geostatistical modeling involves the use of Gaussian processes that are ubiquitous throughout many different fields and are readily extended to the temporal and spatio-temporal contexts (Cressie and Wikle, 2011) as well as commonly employed in computer model emulation (e.g., Higdon et al., 2008) and trajectory estimation (e.g., Hooten and Johnson, 2017).  The use of Gaussian processes in spatially-explicit models has a long history in statistics, but has experienced a resurgence lately due to the need to flexibly and efficiently model large data sets and provide optimal predictions in space and time (Gelfand and Schliep, 2016; Heaton et al., 2019).  

For this example, we specify a version of the full Bayesian geostatistical model (Arab et al., 2017) as 

\begin{align} 
  \mathbf{y} &\sim \text{N}(\mathbf{X}\boldsymbol\beta,\boldsymbol\Sigma(\sigma^2,\phi,\tau^2)) \,  \label{eq:GP1} \\
  \boldsymbol\beta &\sim \text{N}(\boldsymbol\mu_\beta,\boldsymbol\Sigma_\beta) \, \label{eq:GP2} \\
  \sigma^2 &\sim \text{Inv-}\chi^2 (\alpha_1,\alpha_2) \, \label{eq:GP3} \\
  \phi &\sim \text{half-N}(0,\gamma) \, \label{eq:GP4} \\
  \tau^2 &\sim \text{Unif}(0,1)  \, \label{eq:GP5} 
\end{align} 
\noindent where the full data set is denoted as $\mathbf{y}\equiv (y_1,\ldots,y_n)'$ and comprises observations of SST ($^\text{o}$C)  at locations $\mathbf{s}_1,\ldots,\mathbf{s}_n$ in continuous space ${\cal S}$. The spatial covariance of $\mathbf{y}$ is modeled with sill, range, and nugget parameters as
\begin{equation}
\boldsymbol\Sigma(\sigma^2,\phi,\tau^2)\equiv \sigma^2((1-\tau^2)\mathbf{R}(\phi) + \tau^2\mathbf{I})\,.
\end{equation}
We used a Mat\'ern (Mat\'ern, 1986; Guttorp and Gneiting, 2006) covariance function with smoothness parameter set to $3/2$ to model the latent spatial structure and parameterize the correlation matrix $\mathbf{R}(\phi)$. The entries of the correlation matrix are $R_{ij} = (1+d_{ij}/\phi)\exp(-d_{ij}/\phi)$, where $\phi$ is a parameter that controls the range of spatial structure and the Euclidean distance between locations $\mathbf{s}_i$ and $\mathbf{s}_j$ is $d_{ij}=||\mathbf{s}_i-\mathbf{s}_j||_2$. For simplicity, the spatial process is assumed to be second-order stationary and isotropic (although our PP-RB approach can be applied in cases with more general assumptions as well).  For covariates, we used the easting and northing associated with each spatial location.

The full posterior distribution associated with our geostatistical model is $[\boldsymbol\beta,\sigma^2,\phi, \tau^2|\mathbf{y}]\propto [\mathbf{y}|\boldsymbol\beta,\sigma^2,\phi, \tau^2][\boldsymbol\beta][\sigma^2][\phi][\tau^2]$.  To fit the full geostatistical model in (\ref{eq:GP1})--(\ref{eq:GP5}), we constructed a MCMC algorithm based on conjugate updates for $\boldsymbol\beta$ and $\sigma^2$, and used a M-H update for $\phi$ and $\tau^2$ (see Appendix C for details on the implementation). For our example, we used conjugate Jeffreys specifications (Jeffreys, 1946) for the priors $[\boldsymbol{\beta}]\propto 1$ and $[\sigma^2] = 1/\sigma^2$. 

The general PP-RB procedure to fit the Bayesian geostatistical model for $J>3$ partitions of the data involves the following steps:
\begin{enumerate}
  \item Partition the data into $\mathbf{y}\equiv(\mathbf{y}'_1,\ldots,\mathbf{y}'_J)'$ subsets.  
  \item Stage 1: Fit the Bayesian geostatistical model in (\ref{eq:GP1})--(\ref{eq:GP5}) to the first partition of data to yield a MCMC sample from $[\boldsymbol\beta,\sigma^2,\phi,\tau^2|\mathbf{y}_1]$ resulting in realizations $\boldsymbol\beta^{(k)}$, $\sigma^{2(k)}$, $\phi^{(k)}$, $\tau^{2(k)}$ for MCMC iteration $k=1,\ldots,K$.     
  \item Calculate $\log[\mathbf{y}_j|\boldsymbol\beta^{(k)},\sigma^{2(k)},\phi^{(k)},\tau^{2(k)},\mathbf{y}_{1:(j-1)}]$ for realizations $k=1,\ldots,K$ and partitions $j=2,\ldots,J$, in parallel.   
  \item Stage 2: Perform block M-H updates for model parameters using $\boldsymbol\beta^{(k)}$, $\sigma^{2(k)}$, $\phi^{(k)}$, and $\tau^{2(k)}$ randomly from the first stage transient posterior as proposals in (\ref{eq:PP-RBI_MH}) according to Appendix A.  
  \item Stage 3: Sampling randomly from the resulting MCMC sample from the second stage as proposals (Appendix A), perform the third stage M-H updates based on the ratio (\ref{eq:PP-RBI_MH}). 
  \item Stages $4$--$J$: Repeat for all stages, conditioning on the posterior from the previous stage each time.
\end{enumerate}
The precalculation step between stages 1 and 2 in our PP-RB procedure is the computational crux because we must evaluate the log density of the conditional Gaussian distribution repetitively.  Based on well-known multivariate Gaussian properties (e.g., Gentle, 2007), the $j$th conditional data distribution for our geostatistical model is $[\mathbf{y}_j|\boldsymbol\beta^{(k)},\sigma^{2(k)},\phi^{(k)}, \tau^{2(k)}, \mathbf{y}_{1:(j-1)}]=\text{N}(\tilde{\boldsymbol\mu}_j,\tilde{\boldsymbol\Sigma}_j)$, with conditional mean and covariance  

\begin{align}  
  \tilde{\boldsymbol\mu}_j &\equiv \mathbf{X}_j\boldsymbol\beta + \boldsymbol\Sigma_{j,1:(j-1)}\boldsymbol\Sigma_{1:(j-1),1:(j-1)}^{-1}(\mathbf{y}_{1:(j-1)}-\mathbf{X}_{1:(j-1)}\boldsymbol\beta) \,, \label{eq:condmean} \\ 
  \tilde{\boldsymbol\Sigma}_j &\equiv \boldsymbol\Sigma_{j,j}-\boldsymbol\Sigma_{j,1:(j-1)}\boldsymbol\Sigma_{1:(j-1),1:(j-1)}^{-1}\boldsymbol\Sigma_{1:(j-1),j} \,. \label{eq:condvar} 
\end{align}  
\noindent Thus, to evaluate the conditional data model, we must calculate two matrix inverses as well as several matrix products and a determinant.  The floating point operations (FLOPS) associated with inverting $\boldsymbol\Sigma_{1:(j-1),1:(j-1)}$ are the most numerically intensive, on the order of $O(n_{1:(j-1)}^3)$ (where $n_{1:(j-1)}$ is the dimension of $\mathbf{y}_{1:(j-1)}$), which is less than $O(n^3)$ required for the full data set.  In the case where we have two equal sized partitions (i.e., $J=2$), the FLOPS associated with matrix inverses are $O(2(\frac{n}{2})^3)=O(\frac{n^3}{4})$, four times less than for the full data set.  Additionally, after the log conditional data model is evaluated for a given set of parameters, we do not need to retain its mean and covariance matrix, which reduces our storage requirements substantially.        

We applied the PP-RB approach to fit the Bayesian geostatistical model to the SST data using $J=3$ partitions and $K=200000$ MCMC iterations.  Figure~\ref{fig:geo_sst} shows the full spatial data set and the $J=3$ partitions of data, subsampled randomly from the full data set.  The computational time required to perform the entire PP-RB procedure based on a first stage model fit with $K=200000$ MCMC iterations was 1.7 hours whereas the time required to fit the full model with the same number of MCMC iterations was 6.9 hours.  Thus, our PP-RB approach based on $J=3$ random partitions of the spatial data resulted in an algorithm that was approximately 4 times faster to obtain the same inference from the exact model without approximating the covariance function.

\begin{figure}[htp]
  \centering
  \includegraphics[width=1\textwidth]{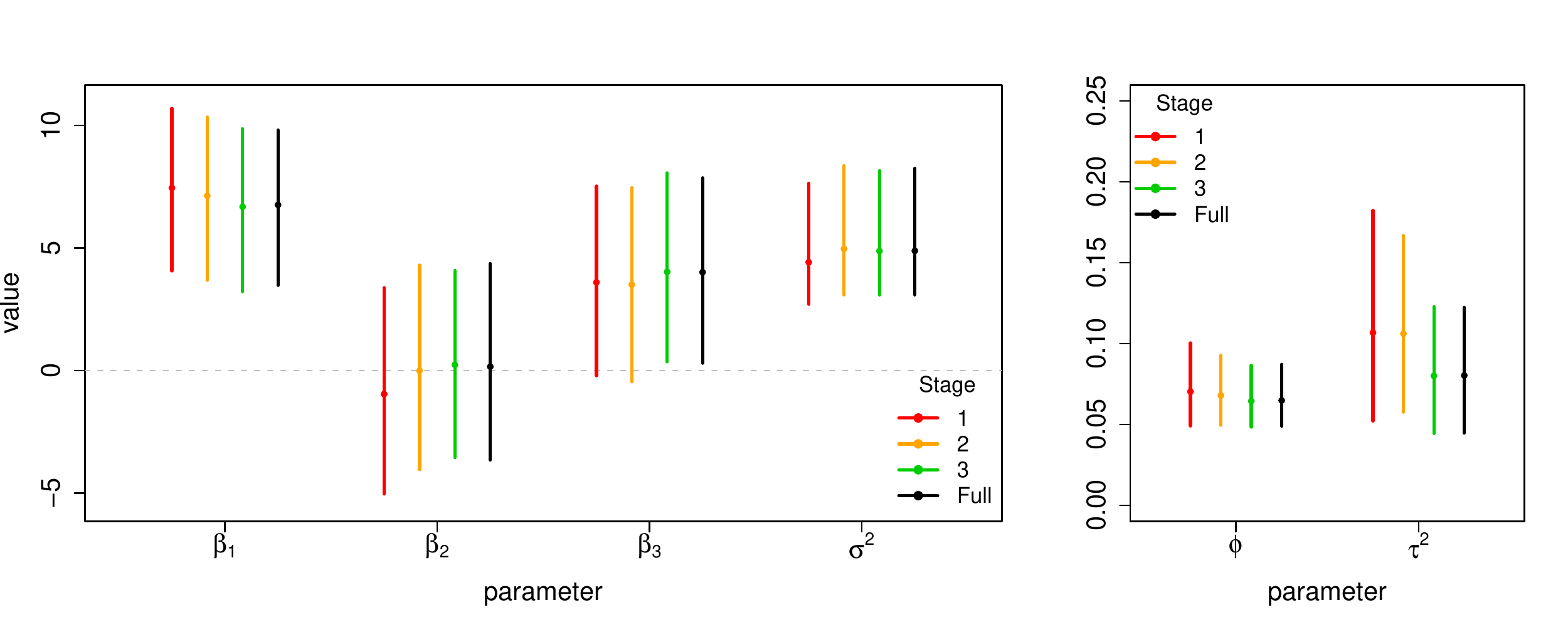}
  \caption{Posterior means (points) and 95\% credible intervals resulting from fitting the Bayesian geostatisical model to the full data set (black) and for each partition $j=1,\ldots,3$ using the PP-RB approach.}
  \label{fig:SST_post}
\end{figure}

We summarized the inference resulting from the two model fits in Figure~\ref{fig:SST_post}, where the 95\% credible intervals and posterior means for each parameter are shown for the full model (in black) and for each stage of the PP-RB procedure in colors ranging from red (stage 1) to green (stage 3).  
It is clear that our inference concerning all geostatistical model parameters improves as we fit the models in each stage of the PP-RB procedure recursively (the green credible intervals match the black ones in Figure~\ref{fig:SST_post}).  In particular, for the $\beta_3$ regression coefficient (associated with the northing covariate), our inference changes from non-significant to significantly different than zero (based on the 95\% credible interval) between the second and third stages as the credible interval is shrunk toward the full-data posterior when we incorporate additional partitions of data. 

\section{PP-RB Application to Online Updating}
To illustrate the PP-RB approach for ``online'' updating (i.e., a strategy for efficiently assimilating new data as they become available; Shifano et al., 2016; Wang et al., 2018), we analyzed temporal data resulting from surveys of Steller sea lion populations.  Steller sea lions are listed as endangered under the U.S. Endangered Species Act over much of their geographic range.  The National Marine Fisheries Service of the U.S. Federal Government monitors the status of this species in Alaska by conducting aerial surveys to count the number of sea lion pups born in the Aleutian Islands and Gulf of Alaska each year.  We focused our analysis on counts at two different rookery sites (Marmot and Sugarloaf) monitored during 1978--2013; although both sites were not surveyed every year and survey effort was generally sparse early in the monitoring program.  The Steller sea lion pup count data are available in the R package `agTrend.'

We analyzed Steller sea lion pup counts using the hierarchical model
\begin{align}
  y_{s,t} & \sim\text{Pois}(\lambda_{s,t}),\label{eq:ssl_obs}\\
  \log(\lambda_{s,1}) & \sim\text{N}(\mu_1,\sigma^2_1)\,, \\
  \log\left(\lambda_{s,t}\right) & \sim\text{N}(\phi_s+\log\left(\lambda_{s,t-1}\right),\sigma_s^2) \,, \\
  \phi_{s} & \sim\text{N}(0,\sigma_\phi^2) \,, \\
  \sigma_s^2 & \sim\text{IG}(\alpha,\beta) \,,
\end{align}
\noindent where $y_{s,t}$ is the observed pup count at sites $s=1,2$ (i.e., Marmot and Sugarloaf sites) in year $t=1,\ldots,T$ (see Appendix D for implementation details).  These sites were not monitored in year $T+1=2014$, thus we sought to use the available data from 1978 through 2013 ($\mathbf{y}_{s,1:T}$) to predict sea lion pup count intensity in the year 2015 ($\lambda_{s,T+2}$) as rapidly as possible after obtaining the subsequent observations $y_{s,T+2}$.  We used the PP-RB approach to accomplish this task without the burden of fitting the model to the full data set. 

In the context of online updating, we assume a first-stage analysis has been conducted based on previous pup counts $\mathbf{y}_{s,1:T}$ resulting in a MCMC sample comprised of $\boldsymbol{\theta}_s^{(k)}\equiv(\phi_s^{(k)},\sigma_s^{2(k)},\boldsymbol\lambda_s^{(k)})'$ (where $k=1,\ldots,K$ indexes MCMC iterations from the first-stage analysis).  When new data $y_{s,T+2}$ arrive and we wish to update inference using the second-stage algorithm, we first sample the new intensity parameters $\lambda_{s,T+1}$ and $\lambda_{s,T+2}$ from their predictive distributions 

\begin{align}
  \log\left(\lambda_{s,T+1}^{(*)}\right)&\sim\text{N}\left(\phi_s^{(k)}+\log\left(\lambda_{s,T}^{(k)}\right),\sigma_s^{2(k)}\right)\,, \\
  \log\left(\lambda_{s,T+2}^{(*)}\right)&\sim\text{N}\left(\phi_s^{(k)}+\log\left(\lambda_{s,T+1}^{(*)}\right),\sigma_s^{2(k)}\right)\,. 
\end{align}
Then, the $k$th M-H acceptance ratio to update all parameters, including $\lambda_{s,T+2}$, in the second-stage analysis is 
\begin{align}
  r&=\frac{[y_{s,T+2}|\lambda_{s,T+2}^{(*)}][\lambda^{(*)}_{s,T+2},\lambda^{(*)}_{s,T+1},\boldsymbol{\theta}_s^{(*)}|\mathbf{y}_{s,1:T}][\lambda^{(k-1)}_{s,T+2},\lambda^{(k-1)}_{s,T+1},\boldsymbol{\theta}_s^{(k-1)}|\mathbf{y}_{s,1:T}]}{[y_{s,T+2}|\lambda_{s,T+2}^{(k-1)}][\lambda^{(k-1)}_{s,T+2},\lambda^{(k-1)}_{s,T+1},\boldsymbol{\theta}_s^{(k-1)}|\mathbf{y}_{s,1:T}][\lambda^{(*)}_{s,T+2},\lambda^{(*)}_{s,T+1},\boldsymbol{\theta}_s^{(*)}|\mathbf{y}_{s,1:T}]} \,, \\
   &=\frac{[y_{s,T+2}|\lambda_{s,T+2}^{(*)}]}{[y_{s,T+2}|\lambda_{s,T+2}^{(k-1)}]} \,, \label{eq:ssl_mh}
\end{align}
\noindent which simplifies to a function containing only Poisson probability mass functions resulting from the fact that the proposal distribution for $\left(\lambda_{s,T+2},\lambda_{s,T+1},\boldsymbol{\theta}_s\right)$ is $\left[\lambda_{s,T+2},\lambda_{s,T+1},\boldsymbol{\theta}_s|\mathbf{y}_{s,1:T}\right]$, which is sampled during the first-stage analysis.  As before, we draw proposals $\lambda_{s,T+2}^{(*)}$ randomly with replacement from the stage one MCMC posterior predictive sample (Appendix A) and accept the proposal with probability $\text{min}(r,1)$.  

\begin{figure}[htp]
\centering \includegraphics[width=3.25in]{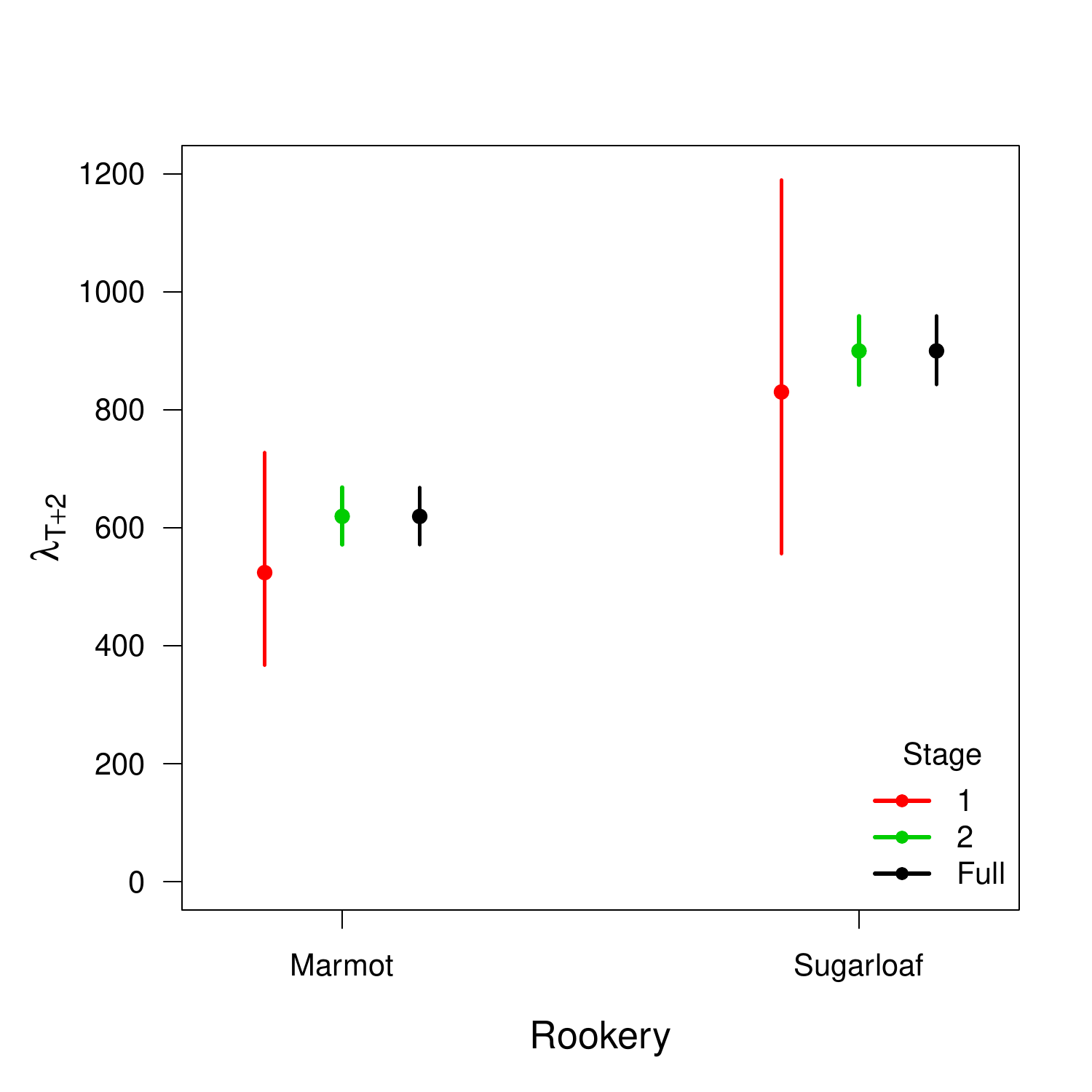}
\caption{Posterior means (points) and 95\% credible intervals (vertical lines) for Steller sea lion pup count intensities $\lambda_{s,T+2}$ at two sites in the Aleutian Islands, AK for year T+2=2015. The black intervals correspond to the results from the full model, whereas the red and green intervals correspond to the inference obtained from the first- and second-stages of the PP-RB analysis.}
\label{fig:ssl_results} 
\end{figure}

To compare the PP-RB method for online updating with the model fit to the entire data set simultaneously, we used $K=100000$ MCMC iterations for both the full data set ($\mathbf{y}_{s,1:T}$ and $y_{s,T+2}$) and the data set without the last year of data ($\mathbf{y}_{s,1:T}$).  We then relied on the PP-RB method to assimilate the final year of data $y_{s,T+2}$ in a second algorithm using the M-H updates described in (\ref{eq:ssl_mh}).  Although predictions from the first-stage analysis were highly uncertain (i.e., wide red credible intervals in Figure~\ref{fig:ssl_results}), upon incorporation of the new data ($y_{s,T+2}$), inference was virtually identical to the full-data posterior (i.e., green and black credible intervals match in Figure~\ref{fig:ssl_results}). 
For this analysis, updating our inference using the second-stage algorithm and the final year of data was 59 times faster than fitting the model to the full data set simultaneously.

\section{Discussion}
In this era where new data are constantly streaming in and both sensing and storage technology are improving, online statistical models have become more challenging to fit efficiently.  Dietz et al. (2018) made a strong case for the need to fit statistical models to incoming data operationally and regularly provide iterative forecasts based on important ecological and environmental data streams.  Statistically rigorous recursive inference and forecasting is clearly useful in all fields, but existing methods for assimilating new data recursively or performing meta-analysis may be inaccessible to practitioners or computationally infeasible.  Our PP-RB approach relies on well-known Bayesian updating principles and commonly used MCMC methods for fitting models.  The PP-RB approach combines two existing RB concepts (i.e., Prior- and Proposal-RB) to result in a broadly applicable multi-stage technique for fitting Bayesian models sequentially to partitioned data sets.  

Overall, the PP-RB method we presented is very accessible to practitioners because it relies on a first-stage posterior sample (that can be acquired using automated software) followed by a sequence of simple M-H updates.  The multicore architecture of modern computers can be leveraged to accelerate the PP-RB by precomputing the conditional log likelihoods for each first-stage MCMC sample in parallel, but parallel computing is not necessary to use PP-RB in general.  For example, when the results of a previous analysis are available and we only seek to obtain inference based on a single new incoming data partition (i.e., partition $\mathbf{y}_J$), no recursion is necessary.  In that case, we simply condition on the most recent model output (i.e., based on partitions $\mathbf{y}_{1:(J-1)}$).  

The main challenge associated with the PP-RB method is that we need to evaluate conditional log likelihoods.  The PP-RB procedure is straightforward when the conditional log likelihood function is analytically tractable (such as for the multivariate normal distribution), but approximation of the conditional log likelihood must be used when it is not tractable.  We have had good success approximating the conditional log likelihood using SMC and, in some cases, pseudo-likelihood approaches in a variety of other applications.   

The data partitions required in PP-RB need not correspond to a meaningful aspect of space, time, or model structure, but in some cases, natural partitions may be available (i.e., spatio-temporal data) and can be used.  In fact, partition design is an important area of future research related to PP-RB because it could lead to optimal recursive strategies and even faster inference.  In fact, Gramacy et al. (2015) and Guinness (2018) explored similar concepts related to the design of partitions for fitting approximate Gaussian process models more efficiently.  Those partitioning concepts can be used in conjunction with our PP-RB approach and may extend to a broader set of Bayesian statistical models.  However, poorly selected partitions may result in suboptimal inference because the early stages could result in MCMC samples that do not adequately explore the correct posterior in practice.  For example, Zimmerman (2006) found that designs with clusters comprised of sampling locations that are near each other in space facilitate the estimation of covariance parameters in geostatistical models.     

Many other approaches to Gaussian process approximations have been developed over the past several decades and are appearing with greater regularity recently.  For example, Vecchia (1988) presented a Gaussian process approximation based on the same type of recursive expression of the data model we used in our geostatistical example from the previous section (also see Stein, 2004; Huang and Sun, 2016; Katzfuss and Guinness, 2018).  This concept led to several recursive approaches to developing approximate Gaussian process models that have been proposed recently, including predictive processes (Banerjee et al., 2008) and nearest neighbor predictive processes (Datta et al., 2016; Finley et al., 2018), both of which are compatible with our PP-RB method.  Furthermore, any of the alternative approaches for approximating Gaussian process covariance matrices using reduced-rank or sparse parameterizations (e.g., Higdon, 2002; Furrer et al., 2006; Cressie and Johannesson, 2008; Wikle, 2010; Lindgren et al., 2011; Gramacy et al., 2015; Nychka et al., 2015; Katzfuss, 2017) are also compatible with our PP-RB method, as long as they are applied in a Bayesian context (also see Heaton et al., 2019 for an excellent review).  Finally, there may be value in pairing subsampling methods (e.g., Liang et al., 2013; Kleiner et al., 2014;  MacLaurin and Adams, 2015; Barbian and Assuncao, 2017) with PP-RB to reduce computational requirements further. 

The natural recursive nature of the PP-RB method is not limited to use for improving computational efficiency, it also reconciles well with optimal design and monitoring strategies.  Optimal adaptive design, especially in the spatio-temporal context, is becoming more popular in environmental (e.g., Wikle and Royle, 1999, 2005) and ecological statistics (e.g., Hooten et al., 2009, 2012; Williams et al., 2018).  Our PP-RB method can be used to rapidly assimilate new data and characterize posterior forecast distributions that can be optimized to reduce the uncertainty associated with ongoing monitoring efforts without requiring a reanalysis of the entire cumulative data set.      

In terms of alternative methods for efficient Bayesian computing, a variety of computing strategies have become popular because of increasing computational demands due to larger data sets and more complex models.  For example, related classes of computing strategies are: Consensus MC (Scott et al., 2016), Weierstrass samplers (Wang and Dunson, 2013), embarrassingly parallel MCMC (Neiswanger et al., 2013), and Modular Bayes (Jacob et al., 2017), among others.    

\section*{References}

\rf Albert, J.H. and S. Chib. (1993). Bayesian analysis of binary and polychotomous response data.  Journal of the American Statistical Association, 88: 669-679.

\rf Altman, M., A. Wood, D.R. O'Brien, and U. Gasser. (2018). Practical approaches to big data privacy over time. International Data Privacy Law, 8: 29-51.

\rf Andrieu, C., A. Doucet, and R. Holenstein. (2010). Particle Markov chain Monte Carlo methods. Journal of the Royal Statistical Society: Series B, 72: 269-342.

\rf Arab, A., M.B. Hooten, and C.K. Wikle. (2017). Hierarchical Spatial Models. In: Encyclopedia of Geographical Information Science, Second Edition. Springer.

\rf Barbian, M.H. and R.M. Assuncao.  (2017). Spatial subsemble estimator for large geostatistical data. Spatial Statistics, 22: 68-88.

\rf Berliner, L. M. 1996. Hierarchical Bayesian time series models. Pages 15-22 in K. Hanson and R. Silver, editors. Maximum entropy and Bayesian methods. Kluwer Academic, Dordrecht, The Netherlands.

\rf Bradford, R., and A. Thomas. (1996). Markov chain Monte Carlo methods for family trees using a parallel processor.  Statistics and Computing, 6: 67-75.

\rf Brockwell, A.E.  (2006).  Parallel Markov chain Monte Carlo simulation by pre-fetching.  Journal of Computational and Graphical Statistics, 15: 246-261.

\rf Chopin, N.  (2002). A sequential particle filter method for static models.  Biometrika, 89: 539-552. 

\rf Chopin, N., P.E. Jacob, and O. Papaspiliopoulos.  (2013).  SMC$^2$: An efficient algorithm for sequential analysis of state space models.  Journal of the Royal Statistical Society: Series B, 75: 397-426.

\rf Cressie, N.A.C.  (1993).  Statistics for Spatial Data, Revised Edition.  Wiley.

\rf Cressie, N.A.C., C.A. Calder, J.S. Clark, J.M. Ver Hoef, and C.K. Wikle. 2009. Accounting for uncertainty in ecological analysis: The strengths and limitations of hierarchical statistical modeling. Ecological Applications, 19: 553-570.

\rf Cressie, N.A.C. and G. Johannesson. (2008). Fixed rank kriging for very large spatial data sets. Journal of the Royal Statistical Society: Series B, 70: 209-226

\rf Cressie, N.A.C. and C.K. Wikle.  (2011).  Statistics for Spatio-Temporal Data.  Wiley.  

\rf Datta, A., S. Banerjee, A.O. Finley, and A.E. Gelfand. (2016). Hierarchical nearest-neighbor Gaussian process models for large geostatistical datasets.  Journal of the American Statistical Association, 111: 800-812.

\rf Dietze, M., A. Fox, L. Beck-Johnson, J.L. Betancourt, M.B. Hooten, C. Jarnevitch, T. Kiett, M. Kenney, C. Laney, L. Larsen, H. Loescher, C. Lunch, B. Pijanowski, J. Randerson, E. Reid, A. Tredennick, R. Vargas, K. Weathers, and E. White. (2018). Iterative near-term ecological forecasting: Needs, opportunities, and challenges. Proceedings of the National Academy of Sciences, In Press.

\rf Del Moral, P., A. Doucet, and A. Jasra.  (2006).  Sequential Monte Carlo samplers.  Journal of the Royal Statistical Society: Series B, 68: 411-436.   

\rf Dold, H.M. and I. Fr\"und. (2014). Joint Bayesian inference reveals model properties shared between multiple experimental conditions. PloS One, 9: e91710.

\rf Doucet, A., J.F.G. de Freitas, and N.J. Gordon. (2001).  Sequential Monte Carlo Methods in Practice.  Springer.  New York. 

\rf Finley, A., A. Datta, B.C. Cook, D.C. Morton, H.E. Andersen, and S. Banerjee.  (2018).  Efficient algorithms for Bayesian nearest neighbor Gaussian processes.  arXiv: arXiv:1702.00434v3.

\rf Furrer, R., M.G. Genton, and D. Nychka. (2006). Covariance tapering for interpolation of large spatial datasets. Journal of Computational and Graphical Statistics, 15: 502-523.

\rf Garrard, G.E., M.A. McCarthy, P.A. Vesk, J.Q. Radford, and A.F. Bennett. (2012).  A predictive model of avian natal dispersal distance provides prior information for investigating response to landscape change. Journal of Animal Ecology, 81: 14-23.

\rf Gelfand, A.E. and S. Ghosh.  (2015).  Hierarchical Modeling.  In: Bayesian Theory and Applications.  Eds:  P. Damien, P. Dellaportas, N.G. Polson, and D.A. Stephens.  Oxford University Press.  

\rf Gelfand, A.E. and A.F. Smith. (1990). Sampling-based approaches to calculating marginal densities. Journal of the American Statistical Association, 85: 398-409.

\rf Gelfand, A.E. and E.M. Schliep. (2016). Spatial statistics and Gaussian processes: A beautiful marriage. Spatial Statistics, 18: 86-104.

\rf Gentle, J.E. (2007).  Matrix Algebra:  Theory, Computations, and Applications in Statistics.  Springer.

\rf Gerber, B.D., M.B. Hooten, C.P. Peck, M.B. Rice, J.H. Gammonley, A.D. Apa, and A.J. Davis. (2018). Accounting for location uncertainty in azimuthal telemetry data improves ecological inference. Movement Ecology, 6: 14. 

\rf Glynn, P.W. and P. Heidelberger. (1992). Analysis of initial transient deletion for parallel steady-state simulations. SIAM Journal on Scientific and Statistical Computing, 13: 904-922.

\rf Gramacy, R.B. and D.W. Apley.  (2015).  Local Gaussian process approximation for large computer experiments.  Journal of Computational and Graphical Statistics, 24: 561-578.

\rf Green, P., K. Latuszynski, M. Pereyra, and C. Robert. (2015). Bayesian computation: A summary of the current state, and samples backwards and forwards. Statistical Computation, 25: 835-862.

\rf Guinness, J. (2018).  Permutation and grouping methods for sharpening Gaussian process approximations.  Technometrics, In Press.

\rf Guttorp, P. and T. Gneiting. (2006). Studies in the history of probability and statistics XLIX: On the Mat\'ern correlation family, Biometrika: 93, 989-995.

\rf Hanks, E.M., M.B. Hooten, and M. Alldredge. (2015). Continuous-time discrete-space models for animal movement. Annals of Applied Statistics, 9: 145-165. 

\rf Heaton, M.J., A. Datta, A.O. Finley, R. Furrer, R. Guhaniyogi, F. Gerber, R.B. Gramacy, D. Hammerling, M. Katzfuss, F. Lindgren, and D.W. Nychka.  (2019). A case study competition among methods for analyzing large spatial data.  Journal Agricultural, Biological, and Environmental Statistics.

\rf Higdon, D. (2002). Space and space-time modeling using process convolutions. In: Quantitative Methods for Current Environmental Issues. Springer.

\rf Higdon, D., J. Gattiker, B. Williams, and M. Rightley. (2008). Computer model calibration using high-dimensional output. Journal of the American Statistical Association, 103: 570-583.

\rf Hobbs, N.T. and M.B. Hooten. (2015). Bayesian Models: A Statistical Primer for Ecologists. Princeton University Press.

\rf Hooten, M.B. and T.J. Hefley. (2019).  Bringing Bayesian Models to Life. Chapman \& Hall/CRC.

\rf Hooten, M.B. and D.S. Johnson. (2017). Basis function models for animal movement. Journal of the American Statistical Association, 112: 578-589.

\rf Hooten, M.B., C.K. Wikle, S.L. Sheriff, and J.W. Rushin.  (2009).  Optimal spatio-temporal hybrid sampling designs for ecological monitoring. Journal of Vegetation Science, 20: 639-649.

\rf Hooten, M.B., Johnson, D.S., Hanks, E.M., and J.H. Lowry. (2010). Agent-based inference for animal movement and selection. Journal of Agricultural, Biological and Environmental Statistics, 15: 523-538.

\rf Hooten, M.B., F.E. Buderman, B.M. Brost, E.M. Hanks, and J.S. Ivan. (2016). Hierarchical animal movement models for population-level inference. Environmetrics, 27: 322-333. 

\rf Hooten, M.B., B.E. Ross, and C.K. Wikle. (2012). Optimal spatio-temporal monitoring designs for characterizing population trends. In: Design and analysis of long-term ecological monitoring studies.  Gitzen, R.A. J.J. Millspaugh, A.B. Cooper, and D.S. Licht (eds.). Cambridge University Press, Cambridge, Massachusetts, USA.

\rf Huang, H. and Y. Sun. (2018). Hierarchical low rank approximation of likelihoods for large spatial datasets.  Journal of Computational and Graphical Statistics, 27: 110-118.

\rf Hurlbert, S.H. (1984). Pseudoreplication and the design of ecological field experiments.  Ecological Monographs, 54: 187-211.

\rf Jacob, P.E., L.M. Murray, C.C. Holmes, and C.P. Robert. (2017). Better together? Statistical learning in models made of modules. arXiv:1708.08719.

\rf Jeffreys, H. (1946).  An invariant form for the prior probability in estimation problems.  Proceedings of the Royal Society of London, Series A, Mathematical and Physical Sciences, 186: 453-461. 

\rf Jerem, P., K.H.S. Jenni-Eiermann, D. McKeegan, D. McCafferty, and R. Nager. (2018). Eye region surface temperature reflects both energy reserves and circulating glucocorticoids in a wild bird. Scientific Reports, 8: 1907.

\rf Kalman R.E. (1960). A new approach to linear filtering and prediction problems. Journal of Basic Engineering. 82: 35-45.

\rf Katzfuss, M. (2017). A multi-resolution approximation for massive spatial datasets.  Journal of the American Statistical Association, 112: 201-214.

\rf Katzfuss, M. and J. Guinness.  (2018).  A general framework for Vecchia approximations of Gaussian processes.  arXiv: 1708.06302v3.

\rf Kleiner A., A. Talwalkar, P. Sarkar, M.I. Jordan. (2014). A Scalable bootstrap for massive data.  Journal of the Royal Statistical Society: Series B, 76: 795-816. 

\rf Liang F., Y. Cheng, Q. Song, J. Park, P.A. Yang. (2013). resampling-based stochastic approximation method for analysis of large geostatistical Data. Journal of the American Statistical Association, 108: 325-339.

\rf Lindgren, F., H. Rue, and J. Lindstrom. (2011).  An explicit link between Gaussian fields and Gaussian Markov random fields: The stochastic partial differential equation approach. Journal of the Royal Statistical Society: Series B, 73: 423-498.

\rf Lunn D., J. Barrett, M. Sweeting, S. Thompson. (2013). Fully {B}ayesian hierarchical modelling in two stages, with application to meta-analysis. Journal of the Royal Statistical Society: Series C (Applied Statistics), 62: 551-572.

\rf D. MacLaurin and R.P. Adams. (2014). Firefly Monte Carlo: Exact MCMC with subsets of data. In Proceedings of the Conference on Uncertainty in Artificial Intelligence.

\rf Martin, J., H.H. Edwards, C.J. Fonnesbeck, S.M. Koslovsky, C.W. Harmak, and T.M. Dane. (2015). Combining information for monitoring at large spatial scales: First statewide abundance estimate of the Florida manatee. Biological Conservation, 186: 44-51.

\rf Mat\'ern, B. (1986), Spatial Variation (2nd ed.), Berlin: Springer-Verlag.

\rf McCarthy, M.A. and P.I.P. Masters. (2005). Profiting from prior information in {B}ayesian analyses of ecological data. Journal of Applied Ecology, 42: 1012-1019.

\rf Neal, R.M. (2011). MCMC Using Hamiltonian Dynamics. In Handbook of Markov Chain Monte Carlo.  Eds. Brooks, S., A. Gelman, G. L. Jones and X.-L. Meng.  CRC Press, New York.
(S. Brooks, A. Gelman, G. L. Jones and X.-L. Meng, eds.) CRC Press, New York.

\rf Neiswanger, W., C. Wang, and E. Xing. (2013). Asymptotically exact, embarrassingly parallel MCMC.  arXiv:1311.4780.

\rf Nychka, D., S. Bandyopadhyay, D. Hammerling, F. Lindgren, and S. Sain. (2015). A multiresolution Gaussian process model for the analysis of large spatial datasets. Journal of Computational and Graphical Statistics 24: 579-599. 
 
\rf Patil G. and C. Rao. (1977). The weighted distributions: a survey of their applications. In Applications of Statistics, P Krishnaiah (ed.). North Holland Publishing Company: Amsterdam, the Netherlands.

\rf Rosenthal, J.S. (2000). Parallel computing and Monte Carlo algorithms. Far East Journal of Theoretical Statistics, 4: 207-236.

\rf S\"arkk\"a, S. (2013). Bayesian Filtering and Smoothing. Cambridge University Press.

\rf Schifano E.D., J. Wu, C. Wang, J. Yan, M.H. Chen.  (2016).  Online updating of statistical inference in the big data setting. Technometrics, 58: 393-403.

\rf Scott, S.L., A.W. Blocker, F.V. Bonassi, H.A. Chipman, E.I. George, and R.E. McCulloch. (2016). Bayes and big data: The consensus Monte Carlo algorithm. International Journal of Management Science and Engineering Management, 11: 78-88.

\rf Stein, M.L., Z. Chi, and L.J. Welty. (2004). Approximating likelihoods for large spatial data sets.  Journal of the Royal Statistical Society: Series B, 66: 275-296.

\rf Vecchia, A. (1988).  Estimation and model identification for continuous spatial processes.  Journal of the Royal Statistical Society, Series B, 50: 297-312.

\rf Wang, C., M.H. Chen, J. Wu, J. Yan, Y. Zhang, and E. Schifano.  (2018). Online updating method with new variables for big data streams.  Canadian Journal of Statistics, 46: 123-146.

\rf Wang, X. and D.B. Dunson. (2013). Parallelizing MCMC via Weierstrass sampler. arXiv:1312.4605.

\rf Wikle, C.K. and J.A. Royle. (1999). Space-time dynamic design of environmental monitoring networks. Journal of Agricultural, Biological and Environmental Statistics, 4: 489-507

\rf Wikle, C.K. and J.A. Royle. (2005). Dynamic design of ecological monitoring networks for non-Gaussian spatio-temporal data. Environmetrics, 16: 507-522.

\rf Wikle, C.K. (2010). Low-rank representations for spatial processes. In: Handbook of Spatial Statistics, eds. A. Gelfand, P. Diggle, M. Fuentes, and P. Guttorp, Boca Raton, FL: Chapman \& Hall/CRC.

\rf Wikle, C.K. and M.B. Hooten (2010). A general science-based framework for nonlinear spatio-temporal dynamical models. Test, 19: 417-451.

\rf Zimmerman, D.L. (2006). Optimal network design for spatial prediction, covariance parameter estimation, and empirical prediction. Environmetrics, 17: 635-652.

\pagebreak
\section*{Appendix A}
The cancellations in the M-H ratios (i.e., equations (\ref{eq:LunnMH}), (\ref{eq:Lunn_MH_temp}), (\ref{eq:PP-RBI_MH}), and (\ref{eq:ssl_mh})) occur when our proposed parameter values independently arise from the proposal distribution.  Lunn et al.\ (2013) suggested sampling the proposals randomly with replacement from the stage one Markov chains to reduce dependence.  To illustrate this concept, we denote the stage one Markov chain values for the parameters in (\ref{eq:LunnMH}) as $\boldsymbol\beta_j^{(*,k)}$ for $k=1,\ldots,K$ MCMC iterations.  Then if we sample our proposal at random from those values, the implied proposal distribution is a categorical distribution on the set of $\boldsymbol\beta_j^{(*,k)}$ with probabilities $[\boldsymbol\beta_j^{(*,k)}|\mathbf{y}_j]/\sum_{\nu=1}^K [\boldsymbol\beta_j^{(*,\nu)}|\mathbf{y}_j]$ for $k=1,\ldots,K$.  Because the denominator sums over the space of $\boldsymbol\beta$ in our proposal, it is a function of the data $f(\mathbf{y}_j)$ only.  Thus, the proposal distribution has the form 
\begin{align}
  \frac{[\boldsymbol\beta_j^{(*,k)}|\mathbf{y}_j]}{\sum_{\nu=1}^K [\boldsymbol\beta_j^{(*,\nu)}|\mathbf{y}_j]} &\propto f(\mathbf{y}_j)[\boldsymbol\beta_j^{(*,k)}|\mathbf{y}_j] \,, \\ 
  &\propto [\boldsymbol\beta_j^{(*,k)}|\mathbf{y}_j] \,, \\
  &\propto [\mathbf{y}_j|\boldsymbol\beta_j^{(*,k)}][\boldsymbol\beta_j^{(*,k)}] \,,
\end{align}
\noindent as required in (\ref{eq:LunnMH}).  Other options could involve thinning the stage one MCMC sample to reduce dependence in the proposed values or randomly permute the stage one MCMC sample (Hooten and Hefley, 2019).  Thinning is the most common way to reduce dependence in the MCMC sample before making inference using Monte Carlo integration, but it may reduce the number of possible proposal values in Proposal-RB and PP-RB substantially.  By contrast, permuting the stage one MCMC sample will not remove the dependence in the Markov chains completely, but does allow us to use the entire set of potential proposals from the first stage (unlike the sampling with replacement approach).   

\section*{Appendix B}
The full BHM for the blue tit eye region temperature data ($y_{ij}$) is specified as 

\begin{align}
  y_{ij} \sim \text{N}(\mu_j,\sigma^2_j) \,, \\
  \mu_j \sim \text{N}(\mu,\sigma^2) \,, \\
  \sigma^2_j \sim \text{IG}(\alpha,\beta) \,, \\
  \mu \sim \text{N}(\mu_{0},\sigma^2_{0}) \,, \\
  \sigma^2 \sim \text{IG}(\alpha_{0},\beta_{0}) \,,
\end{align}
\noindent for $j=1,\ldots,J$ and $i=1,\ldots,n_j$.  For hyperparameters, we specified $\alpha=0.001$, $\beta=1000$, $\mu_0=0$, $\sigma^2_0=10000$, $\alpha_0=0.001$, and $\beta_0=1000$.   

We fit this full model using a MCMC algorithm and $K=50000$ iterations with conjugate updates for model parameters $\mu_j$, $\sigma^2_j$, $\mu$, and $\sigma^2$.  The full-conditional distributions for model parameters are as follows.  For $\mu_j$ the full-conditional is  $[\mu_j|\cdot]=\text{N}(a^{-1}b,a^{-1})$, with

\begin{align}
  a&\equiv\frac{n_j}{\sigma^2_j}+ \frac{1}{\sigma^2} \;, \\
  b&\equiv\frac{\sum_{i=1}^{n_j}y_{ij}}{\sigma^2_j}+\frac{\mu}{\sigma^2} \;.
\end{align}

For $\sigma^2_j$ the full-conditional is $[\sigma^2_j|\cdot]=\text{IG}(\tilde{\alpha},\tilde{\beta})$, with 

\begin{align}
  \tilde{\alpha}&\equiv\frac{n_j}{2}+\alpha \;, \\
  \tilde{\beta}&\equiv\left(\frac{\sum_{i=1}^{n_j}(y_{ij}-\mu_j)^2}{2}+\frac{1}{\beta}\right)^{-1} \;.
\end{align}

For $\mu$ the full-conditional is $[\mu|\cdot]=\text{N}(a^{-1}b,a^{-1})$, with

\begin{align}
  a&\equiv \frac{J}{\sigma^2}+ \frac{1}{\sigma^2_0} \;, \\
  b&\equiv \frac{\sum_{j=1}^J\mu_j}{\sigma^2}+\frac{\mu_0}{\sigma^2_0} \;.
\end{align}

For $\sigma^2$ the full-conditional is $[\sigma^2|\cdot]=\text{IG}(\tilde{\alpha},\tilde{\beta})$, with 

\begin{align}
  \tilde{\alpha}&\equiv\frac{J}{2}+\alpha \;, \\
  \tilde{\beta}&\equiv\left(\frac{\sum_{j=1}^{J}(\mu_j-\mu)^2}{2}+\frac{1}{\beta}\right)^{-1} \;.
\end{align}

For the Proposal-RB implementation of the blue tit eye region temperature HBM, we used  

\begin{align}
  \mu_j \sim \text{N}(0,10000) \,, \\
  \sigma^2_j \sim \text{IG}(0.001,1000) \,, 
\end{align}
for transient priors in the first stage analyses. 

For the Proposal-RB implementation, we fit the first-stage model using $K=100000$ and thinned the resulting Markov chains to yield $K=10000$ for the second-stage analysis.    

\section*{Appendix C}
The Bayesian geostatistical model for the full data set $\mathbf{y}$ was specified as 

\begin{align} 
  \mathbf{y} &\sim \text{N}(\mathbf{X}\boldsymbol\beta,\boldsymbol\Sigma(\sigma^2,\phi,\tau^2)) \, \\
  \boldsymbol\beta &\sim \text{N}(\boldsymbol\mu_\beta,\boldsymbol\Sigma_\beta) \, \\
  \sigma^2 &\sim \text{Inv-}\chi^2(\alpha_1,\alpha_2) \,  \\
  \phi &\sim \text{half-N}(\gamma^2) \\
  \tau^2 &\sim \text{Unif}(0,1) 
\end{align} 

We fit the Bayesian geostatistical model to the full data set using $K=20000$ MCMC iterations and hyperparameters $\boldsymbol\mu_\beta=(0,0,0)'$, $\boldsymbol\Sigma_\beta^{-1}=\mathbf{0}$ (i.e., flat prior), $\alpha_1=0$, $\alpha_2=0$, and $\gamma=0.05$. The coordinates for the spatial locations were scaled to be within $[0,1]\times [0,1]$ and the $\gamma=0.05$ choice implies that $\approx 95$\% of the posterior mass for the effective range of spatial correlation lies between 0 and 1/3 the maximum distance between spatial locations.

The full-conditional distributions for this geostatistical model are conjugate for $\boldsymbol\beta$ and $\sigma^2$.  For $\boldsymbol\beta$ the full-conditional distribution is $[\boldsymbol\beta|\cdot]=\text{N}(\mathbf{A}^{-1}\mathbf{b},\mathbf{A}^{-1})$ with 

\begin{align}
  \mathbf{A}&\equiv \mathbf{X}'\boldsymbol\Sigma^{-1}\mathbf{X}+\boldsymbol\Sigma^{-1}_\beta   \;, \\
  \mathbf{b}&\equiv \mathbf{y}'\boldsymbol\Sigma^{-1}\mathbf{X}+\boldsymbol\mu'_\beta\boldsymbol\Sigma^{-1}_\beta \;. 
\end{align}  

For $\sigma^2$ the full-conditional distribution is $[\sigma^2|\cdot]=\text{Inv-}\chi^2(\tilde{\alpha}_1,\tilde{\alpha}_2)$ with 

\begin{align}
  \tilde{\alpha}_1&\equiv n+\alpha_1 \;, \\ 
  \tilde{\alpha}_2&\equiv  \frac{\alpha_1 \alpha_2 + nS^2}{\alpha_1 + n}  \;.
\end{align}
where $S^2 = (\mathbf{y}-\mathbf{X}\boldsymbol{\beta})'((1-\tau^2)\mathbf{R}(\phi) +\tau^2\mathbf{I})^{-1} (\mathbf{y}-\mathbf{X}\boldsymbol{\beta})/n$

The full-conditional distribution for the spatial parameters $\phi$ and $\tau^2$ will not be conjugate, but we can sample it using an M-H update in the first-stage algorithm.  We write the full-conditional distribution for $\phi$ and $\tau^2$ as

\begin{equation}   
  [\phi,\tau^2|\cdot]\propto [\mathbf{y}|\boldsymbol\beta,\sigma^2,\phi,\tau^2][\phi][\tau^2] \;.
\end{equation}   
\noindent We use the random walk method with rejection sampling for proposing values of ($\phi^{(*)}, \tau^{2(*)})$ (where we reject the update when $\phi^{(*)}$ or $\tau^2\leq0$ and $\tau^2>1$), the resulting M-H ratio is 
\begin{equation}
  r=\frac{\text{N}(\mathbf{y}|\mathbf{X}\boldsymbol\beta^{(k)},\boldsymbol\Sigma(\sigma^{2(k)},\phi^{(*)}, \tau^{2(*)}))\text{half-N}(\phi^{(*)}|\gamma)}{\text{N}(\mathbf{y}|\mathbf{X}\boldsymbol\beta^{(k)},\boldsymbol\Sigma(\sigma^{2(k)},\phi^{(k-1)},\tau^{2(k-1)})\text{half-N}(\phi^{(k-1)}|\gamma)} \;.
\end{equation}
The random walk proposal distribution is adaptively tuned to reach an acceptance rate of $\approx$ 0.3. 

\section*{Appendix D}
The full BHM for the Steller sea lion count data ($y_{s,t}$) is specified as 
\begin{align}
  y_{s,t} & \sim\text{Pois}(\lambda_{s,t})\,, \\
  \log(\lambda_{s,1}) & \sim\text{N}(\mu_1,\sigma^2_1)\,, \\
  \log\left(\lambda_{s,t}\right) & \sim\text{N}(\phi_s+\log\left(\lambda_{s,t-1}\right),\sigma_s^2) \,, \\
  \phi_{s} & \sim\text{N}(0,\sigma_\phi^2) \,, \\
  \sigma_s^2 & \sim\text{IG}(\alpha,\beta) \,,
\end{align}
\noindent for $s=1,\ldots,2$ and $t=2,\ldots,T$.  For hyperparameters, we specified $\mu_1=8.7$, $\sigma^2_1=1.69$, $\sigma^2_\phi=1$, $\alpha=1$, and $\beta=20$.   

We fit this full model using a MCMC algorithm and $K=100000$ iterations with conjugate updates for model parameters $\phi_s$ and $\sigma^2_s$, and M-H updates for $\lambda_{s,t}$ for all $s$ and $t$.  The full-conditional distributions for model parameters are as follows.  For $\phi_s$, the full-conditional is  $[\phi_s|\cdot]=\text{N}(a^{-1}b,a^{-1})$, with

\begin{align}
  a&\equiv\frac{T-1}{\sigma^2_s}+ \frac{1}{\sigma^2_\phi} \,, \\
  b&\equiv\frac{\sum_{t=2}^{T}(\log(\lambda_{s,t})-\log(\lambda_{s,t-1}))}{\sigma^2_s} \,.
\end{align}

For $\sigma^2_s$ the full-conditional is $[\sigma^2_s|\cdot]=\text{IG}(\tilde{\alpha},\tilde{\beta})$, with 

\begin{align}
  \tilde{\alpha}&\equiv\frac{T-1}{2}+\alpha \;, \\
  \tilde{\beta}&\equiv\left(\frac{\sum_{t=2}^{T}(\lambda_{s,t}-\phi_s-\lambda_{s,t-1})^2}{2}+\frac{1}{\beta}\right)^{-1} \;.
\end{align}

For $\log(\lambda_{s,t})$ the full-conditionals are as follows.  For $t=1$, the full-conditional is
\begin{equation}
  [\log(\lambda_{s,1})|\cdot] \propto [y_{s,1}|\lambda_{s,1}][\log(\lambda_{s,2})|\phi_s,\sigma^2_s,\log(\lambda_{s,1})][\log(\lambda_{s,1})] \,,
\end{equation}
\noindent for $t=2,\ldots,T-1$, the full-conditional is
\begin{equation}
  [\log(\lambda_{s,t})|\cdot] \propto [y_{s,t}|\lambda_{s,t}][\log(\lambda_{s,t+1})|\phi_s,\sigma^2_s,\log(\lambda_{s,t})][\log(\lambda_{s,t})|\phi_s,\sigma^2_s,\log(\lambda_{s,t-1})] \,,
\end{equation}
\noindent and, for $t=T$, the full-conditional is 
\begin{equation}
  [\log(\lambda_{s,T})|\cdot] \propto [y_{s,T}|\lambda_{s,T}][\log(\lambda_{s,T})|\phi_s,\sigma^2_s,\log(\lambda_{s,T-1})] \,.
\end{equation}
We used random walk Metropolis proposals for the log intensity state variables such that $\log(\lambda_{s,t})^{(*)} \sim \text{N}(\log(\lambda_{s,t})^{(k-1)},\sigma^2_\text{tune})$.

\end{document}